\documentclass[useAMS,usenatbib,referee]{mn2e}
\usepackage{graphicx}

\newcommand{\ACrH}{$A~^6\Sigma^+$~}
\newcommand{\XCrH}{$X~^6\Sigma^+$~}

\newcommand{\qvv}{$q_{v'v''}$~}
\newcommand{\AMgH}{$A~^2\Pi$~}
\newcommand{\XMgH}{$X~^2\Sigma^+$~}
\newcommand{\Martin}{Mart\'{i}n~}
\newcommand{\Magazzu}{Magazz\'{u}~}
\newcommand{\Bejar}{B\'{e}jar~}

\newcommand{\MJ}{$M_{\rm J}$~}

\newcommand{\mum}{$\mu$m~}

\newcommand{\HHO}{H$_2$O~}

\newcommand{\Tef}{\mbox{$T_{\rm eff}~$}}
\newcommand{\Msun}{\mbox{\,M$_\odot$~}}

\begin{document}

\title[Bands of CrD and MgD and the ``deuterium test'']
{The electronic bands of CrD, CrH, MgD and MgH: application to the ``deuterium test''}

\author[Ya. V. Pavlenko et al.]{Ya. V. Pavlenko$^{1,2,}$\thanks
{E-mail:yp@mao.kiev.ua}, 
G. J. Harris$^3$,
J. Tennyson$^{3}$, 
H. R. A. Jones$^1$,
\newauthor 
J. M. Brown$^4$,
C. Hill$^4$ and
L.A.Yakovina$^1$\\
$^{1}$Main Astronomical Observatory, Academy of Sciences of Ukraine, 
Golosiiv Woods, Kyiv-127, 03680 Ukraine \\
$^{2}$Centre for Astrophysics Research, University of Hertfordshire,
College Lane, Hatfield, Hertfordshire AL10 9AB, UK\\
$^3$ Department of Physics and Astronomy, University College London, 
      Gower Street, London WC1E 6BT UK \\
$^4$ Oxford University,  Physical \& Theoretical Chemistry Laboratory, South Parks Rd, Oxford OX1 3QZ, UK \\
}

\date{}
\pagerange{\pageref{firstpage}--\pageref{lastpage}} \pubyear{2002}

\maketitle

\label{firstpage}

\begin{abstract}

 We compute opacities for the electronic molecular
 band systems \ACrH -- \XCrH of CrH and CrD, and \AMgH -- \XMgH of MgH and 
MgD. The opacities are computed by making use of existing spectroscopic 
constants for MgH and CrH. These constants are adjusted for the
different reduced masses of MgD and CrD. Frank-Condon factors are used
to provide intensities for the individual vibronic bands. These results are used in the computation of synthetic spectra between \Tef = 1800 and 1200 K with an emphasis on the 
realisation of ``deuterium test'', first proposed by Bejar et al. (1999) 
to distinguish brown dwarfs from planetary mass objects.
We discuss the possible use of CrD and MgD electronic bands for the
``deuterium test". We find CrD to be the more promising of the two deuterides,
potentially, the most useful bands 
of CrH/CrD are the  $\Delta v = +1$  and $\Delta v = -1$ at 0.795 and 0.968 \mum .

\end{abstract}

\section{Introduction}

   The ``deuterium test'' was suggested as a method of
identifying planetary mass objects among cool objects (\Bejar et al. 1999,
Chabrier et al. 2000). In practise, it was proposed to search for absorption lines
of molecules containing deuterium (HDO, CrD, FeD,
etc.).
Deuterium is burnt in stellar interiors via
the fusion reaction $^2$D(p,$\gamma$)$^3$He at  temperatures ($T >
8 \times 10^5$~ K). The interiors of substellar 
objects with M $<$ 13~ \MJ, where \MJ is a Jovian mass (0.001 \Msun), 
do not reach temperatures high enough for 
deuterium to ignite (Saumon et al. 1996). As a 
result the deuterium abundance in the atmospheres of these objects 
is unchanged from the formation of these objects.
This gives rise to the definition of a brown dwarf as an object
which has insufficient mass to fuse $^1$H to $^4$He, but has sufficient
mass to fuse $^2$H to $^3$He.  By comparison, a planetary mass object has
insufficient mass to ignite fusion of any sort (Saumon et al. 1996).

In higher-mass objects such as stars,
deuterium burning is completed comparatively quickly (t = 1
-- 3 million years) during the evolution prior to the star's
period on the main sequence (D'Antona \& Mazitelli 1998).
The deuterium depletion rate depends on the mass of the
star or brown dwarf and thus the ``deuterium test'' can be used in a 
number of ways.
\begin{itemize}
\item Discern planets from brown dwarfs in a population of low-mass objects.

\item Determine the evolutionary status of young objects in
open clusters with ages of several million years.
 
\item Study the evolution of the abundance of deuterium
in the atmospheres of low-mass substellar objects, a phenomenon which
is poorly understood. The rate of  depletion of deuterium 
depends upon rotation, magnetic field strength,
and other parameters which affect the efficiency of convection in low mass 
objects. 
\end{itemize}

Such investigations can usefully be combined with
  the ``lithium test''
  proposed by Rebolo et al. (1992), \Magazzu et al. (1993).
 For stars (M $>$ 75 \MJ), the burning of lithium, Li (p,$\alpha) ^4$He,
becomes efficient at evolutionary stages preceding
  the main sequence at interior temperatures of T $>$
  $2.5\times 10^6$ K (D'Antona \& Mazitelli 1998).
The ``lithium test'' has been successfully applied to identify brown dwarfs
in a population of ultracool dwarf stars (Rebolo et al. 1996, Basri
2000). The lithium test is relatively easily applied to M-dwarfs.
The reason for this is that the resonance lines of
neutral atomic lithium lie in the optical region of the spectrum, at 0.6708
\mum. One blemish with the ``lithium test'' is the severe blending of lithium lines with
the background of TiO lines, nonetheless Li lines break through the molecular
background. In the spectra of cooler M and L dwarfs Ti atoms are
depleted onto dust particles (Tsuji et al. 1996, Jones \&
Tsuji 1997, Pavlenko 1998) so TiO absorption is weakened.
In the L dwarf regime the appearance of lithium lines contends 
with dust opacity and the wings of K I and Na I resonance lines 
(see Pavlenko et al. 2000 and references therein).

The realisation of the ``deuterium test'' is considerably more difficult.
Conventionally, the deuterium abundance of hot objects is
determined from analysis of multicomponent features on the
background of the $L_\alpha$ 0.1215 \mum ~H~I line seen in emission.
Unfortunately, this method cannot be used in the case of
ultracool dwarfs which are covered by a thick envelope of neutral hydrogen.
Observations also yield 
H~I emission lines ($H_\alpha$ 0.6563 \mum)
formed in the
outermost layers of hot chromospheres or accretion disks around
young stars.  However, such a plasma is variable and may be polluted by  interstellar material.

Taking into account these circumstances, it is logical to analyse the
spectral lines of deuterated molecules formed in the photospheric
layers of ultracool dwarfs. The first investigation and analysis of
the combined spectra of \HHO /HDO were carried out by Chabrier et
al. (2000) and Pavlenko (2002). Due to the change in mass and the 
breakdown of molecular symmetry the vibration-rotation bands of 
HDO in the mid-infrared spectrum shift with respect to the
\HHO bands. There are a few spectral regions which can be used for
detection of HDO lines in the IR spectra of ultracool dwarfs: 3.5
-- 4, $\sim$5, 6-7 \mum (Pavlenko 2002). The main problem is that
 despite the shift in wavelength such HDO lines will be on a background 
of far stronger \HHO lines.

A possible alternative to HDO are the diatomic hydrides.
Strong molecular bands of diatomic metal hydrides
such as MgH and CrH can be observed in the optical spectrum of 
ultracool dwarf stars. The MgH band system \AMgH - \XMgH can be
observed at 0.47--0.6 \mum, and the CrH band system 
\ACrH -- \XCrH\  show absorption features at 0.6--1.5 \mum.

The molecules MgH and CrH have been known in astrophysics for a long time.
MgH has been more extensively studied than CrH, because it can be observed 
in the spectra of G to M type stars. The dissociation
energy of MgH is very low (1.285 eV) so lines of this molecule are very
sensitive to the temperature and gravity variations in stellar
atmospheres. MgH lines were used to determine 
temperatures in the atmospheres of cool giants (Wyller 1961) and the
Sun (Sinha et al. 1979, Sinha \& Joshi 1982), and for the determination of
the surface gravity of stars (Bell \& Gustaffson 1981, Bell at al. 1985,
Berdyugina \& Savanov 1992, Bonnel \& Bell 1993).

The pure rotational spectrum of MgH and MgD radicals (\XMgH) in
their ground state $v$=0 and $v$=1 vibrational modes has been
studied by Ziurys et al. (1993). The first MgH linelist was computed by Kurucz (1993).
Recently, more extensive studies of MgH transitions were performed by
Weck et al. (2003, 2003a, 2003b) and Skory et al. (2003).

Although CrH has been known since Gaydon \& Pearse (1937), 
its electronic spectrum remained relatively unstudied for many years.
Engvold et al. (1980) identified lines of CrH in a spectrum of
sunspots as formed by \XCrH -- \XCrH transitions. They used
the results of studies of multiplicity of $\Sigma$-terms of CrH by
Klehman \& Uhler (1959) and O'Connor (1967). Later Ram,
Jarman \& Bernath (1993) performed a rotational analysis of 0--0
band of the \ACrH -- \XCrH electronic transition and obtained improved rotational constants for the $v'$ =0 vibrational state. 
Combining these results 
with those of Bauschlicher et al. (2001) and Lipus et al. (1991) for
the vibrationally excited transitions, Burrows et al.(2002)
computed an extended linelist for CrH.

Recently Shin et al. (2005) have measured radiative lifetimes of the
$v$ = 0,1 levels of \ACrH state of CrH. These measured lifetimes
are about 16\% -- 45 \% longer that those obtained by Burrows at al. (2002).
These results provide evidence that the oscillator strengths of Burrows et al. (2002)
should be corrected by a factor of 0.8 for at least the transitions to 
the $v'$ = 0.

The submillimeter spectra of CrH and CrD formed by pure
rotational transitions in the ground electronic
state, have been observed in the laboratory by Halfen \& Ziurys (2004).
Electronic bands of MgD and CrD are likely to be located in the same
spectral regions as the corresponding bands of MgH and CrH. In this
paper we model the bands of these molecules to analyse the possibility
of their use for the determination of the D/H ratio in the
atmospheres of ultracool dwarfs.

In section 2 we present a description of the procedures used
to compute the molecular bands of CrH, CrD, MgH and MgD. 
In section 3 we present the vibrational-rotational 
constants of MgH, MgD, CrH and
CrD. In section 4 we present the results of the computation of molecular bands. 
In section 5 we discuss the possibility of using the
electronic bands of diatomic molecules for a
realisation of the ``deuterium test''.

\section{Procedure of computations}

\subsection{Profile of electron bands of diatomic molecule}

In this paper we compute the profile of the electronic band 
averaged over its rotational structure (Naersisian et al. 1989):

\begin{eqnarray}
 \int k_{\omega}d{\omega} & = & \frac{A\  q_{v'v''}\   
S_e S_{j'j''}\  \omega_{j'j''}}{Q(T)} \\ \nonumber
           & & \times \exp \left(-\frac{hc(E_e^{''}+E_v^{''} + E_j^{''})}{kT}\right) \\ \nonumber
&& \times \left[ 1-\exp\left(-\frac{h \omega_{j'j''}}{kT}\right) \right],\nonumber
\end{eqnarray}
where as symbols $''$ label lower levels of transitions, $A = 8
\pi^3(3hc)^{-1}$, 
 $k$ is Boltzmann's 
constant and
$T$ is the temperature,
$\omega$ is a
frequency and $q_{v'v''}$  are Franck-Condon factors for the
corresponding transition, the strength of transition $S_e =
S_e(0,0)\times \omega_{v'v''}/\omega_{0,0}$ (Schadee 1968), $Q(T)$
is the partition function of the molecule.

For a given frequency the sum of the contribution from the $P$, $Q$ and
$R$ branches is:

\begin{eqnarray}
k_{\omega} = \frac{A\ S_e\  F_e}{n \lambda Q(T)} 
\sum q_{v'v''}\  C_{vj} \sum |\lambda_j^v|,
 \end{eqnarray}
where $F_e = \exp (-hcE_e^{''}/(kT))/\delta E_e$, \\
$C_{vj} =(\delta E_e + \delta E_v) (2 j'' +1)
\exp (-hc(E_j^{''}+E_v^{''})) $, \\
$\lambda = 1/ \delta \omega$.

This method of the computation of the profiles of molecular bands was
successfully applied to the modelling of the bands of TiO in the spectra of
ultracool dwarfs (Pavlenko 1997a, 1998a,  Pavlenko
2000), and CN and C$_2$ bands in the spectra of evolved stars (Pavlenko et
al. 2000a,c). It is worth noting that direct comparison of
our JOLA synthetic spectra with more sophisticated ``line by
line'' computations for linelists of a few molecules show good
agreement (Pavlenko 1998a) both in the profiles of bands and positions 
of the band heads.

The JOLA approach has been used for the modelling of the CrH 
bands  $\Delta v$ =0 band system located at 0.8640 \mum
in the spectra of
ultracool dwarfs (Pavlenko 1999, Pavlenko et al. 2000). 
These bands are not
considered for the deuterium test
here, because the wavelengths of corresponding lines of CrH and CrD
are nearly identical.

\subsection{Computations of molecular constants of isotopic molecules}

For a given electronic state, the energy levels of a diatomic molecule 
can, in general, be fitted by:
\begin{eqnarray}
E_{e,v,N} & = & T_e+\omega_e(v+1/2)-\omega_ex_e(v+1/2)^2 \nonumber \\
       &   & +B_eN(N+1)-\alpha_eN(N+1)(v+1/2) \nonumber \\
       &   & -D_e(N(N+1))^2 \nonumber \\
       &   & +\beta_e(N(N+1))^2(v+1/2),
\end{eqnarray}
where $T_e$, $\omega_e$, $\omega_ex_e$, $B_e$, $\alpha_e$, $D_e$ and $\beta_e$ are the
various spectroscopic constants, 
 $v$ and $N$ are vibrational excitation and rotational quantum numbers,
respectively.
To a good approximation the electronic structure of the various 
isotopic analogues of a given molecule are the same. However,
as the nuclei have different masses, the reduced mass and moment 
of inertia of the molecule are different, resulting in different 
rotation-vibration frequencies.
For the case of an anharmonic oscillator and non-rigid rotator model 
of the molecule, the rotational-vibrational constants depend upon the ratio of
reduced mass $\mu_{ab} = m_a \times m_b /(m_a + m_b)$ of two
isotopic molecules (see Wang \& Xia 1996, Craybeal 1988).

The ratios of spectroscopic constants of two isotopic species, labelled
as ``o'' and ``i'', of reduced masses $\mu_o$ and $\mu_i$, depend
on the ratio $\rho =  (\mu_o/\mu_i)^{(1/2)}$.  So that the following approximations
can be used:

$$ (\omega_e)_i/(\omega_e)_o = \rho $$

$$ (\omega_ex_e)_i/(\omega_ex_e)_o = \rho^2 $$

$$ (B_e)_i/(B_e)_o  = \rho^2 $$

$$  (D_e)_i/(D_e)_o  = \rho^4 $$

$$ (\alpha_e)_i/(\alpha_e)_o = \rho^3 $$

$$ (\beta_e)_i/(\beta_e)_o = \rho^5 $$

\begin{table}
 \centering
\caption{\label{_t1} Reduced masses and ratios of reduced masses
for molecules of interest.}
\begin{tabular}{lll}
\hline
\noalign{\smallskip}
Molecule  &  $\mu$/u          &  $\rho$  \\
          &                 &            \\
MgH       &  0.967481    &            \\
          &                 & 0.721116  \\
MgD       & 1.86051      &            \\
          &                 &            \\
CrH       &  0.988438    &            \\
          &                 &  0.713873  \\
CrD       &  1.93958     &            \\
          &                 &            \\
\hline
\noalign{\smallskip}
\end{tabular}
\centering
\end{table}

The reduced masses of the molecules of interest here are given in
Table \ref{_t1} using the dominant species $^{24}$Mg and $^{52}$Cr.
The spectroscopic constants of MgH and CrH have been taken from 
Balfour \&  Cartwright (1976), Lemoine et al (1988) and 
Bauschlicher et al. (2001).
The computed spectroscopic constants for MgD and CrD are
given in Table \ref{_t2}. We adopted oscillators strengths f$_e$
= 0.059 (Kuznezova et al. 1980) and 0.001 (Pavlenko 1999) for the MgH 
(\AMgH - \XMgH) and the CrH (\ACrH - \XCrH) band system, respectively.

We  calculated the Franck-Condon factors \qvv for the  
MgH and CrH band profile computations. The RADEN
program of Kuzmenko et al. (1984), and the FRANKQ 
program (Tsymbal, 1984) were used,
respectively. In the FRANKQ program a Morse potential is used to
represent the true potential, and is solved for rotation-vibration motion. 
We compared the computed \qvv values with the  results of Nicholls (1981) 
and found only small differences. It is worth noting that the 
adopted system of \qvv determines the relative strength of the ($v''$,$v'$)
bands, but not their location in the spectrum. In this work we are
interested in the computation of the main bands of molecules of
interest, therefore the choice of any \qvv does not 
critically affect our main results.


\begin{table*}
\centering \caption{\label{_t2} Spectroscopic constants of
MgH, MgD, CrH and CrD in cm$^{-1}$.}
\begin{tabular}{lllllllll}
\hline
\noalign{\smallskip}
Molecule,        &  &                 &            \\
data source &  &  $T_e$  &  $\omega_e$ & $\omega_ex_e$ & $B_e$ & $\alpha_e$ & $D_e$ & $\beta_e$ \\
\hline
\noalign{\smallskip}
                               &  &                 &            \\
MgH,      &  &          &         &        &         &     &&   \\
Balfour \&  Cartwright (1976) & \AMgH & 19216.8  & 1599.50 & 32.536 & 6.1913 & 0.1931   & 3.60$\times10^{-4}$ & 6.1$\times10^{-6}$ \\
Lemoine et al. (1988)         & \XMgH & 0.00     & 1495.25 & 31.637 & 5.6443 & 0.1845   & 3.53$\times10^{-4}$ & 2.57$\times10^{-6}$\\
MgD,    &     &          &         &        &         &        \\
This work                     & \AMgH & 19216.8  & 1153.43 & 16.918 & 3.2195 & 0.07241  & 9.73$\times10^{-5}$ & 1.2$\times10^{-6}$ \\
Lemoine et al. (1988)         & \XMgH & 0.0      & 1078.14 & 16.147 & 2.9668 & 0.068294 & 9.61$\times10^{-5}$ & 2.79$\times10^{-7}$ \\
    &     &          &          &        &        &         \\
CrH,      &     &          &          &        &        &         \\
Bauschlicher et al. (2001)    & \ACrH &  11552.7 & 1524.80 & 22.280 & 5.3427 & 0.141349 & 3.01$\times10^{-4}$ & 9.18$\times10^{-5}$ \\
                              & \XCrH &  0.0     & 1656.05 & 30.491 & 6.2222 & 0.180978 & 3.52$\times10^{-4}$ & 5.21$\times10^{-6}$ \\
CrD,  & &         &           &       &          &          \\                                                                      
this work                     & \ACrH &  11552.7 & 1088.32 & 11.35  & 2.7227 & 0.051395 & 7.82$\times10^{-5}$ & 1.70$\times10^{-5}$ \\
                              & \XCrH &  0.0     & 1182.00 & 15.533 & 3.1709 & 0.065804 & 9.14$\times10^{-5}$ & 9.66$\times10^{-7}$ \\
                                                                                                                                    \\
Ram \& Bernath (1995)         & \ACrH &  11559.7 & 1066.42 & -      &  2.7627& 0.049687 & 7.03$\times10^{-5}$ &  -                  \\
                              & \XCrH &  11559.7 & 1183.19 & 15.60  & 3.1754 & 0.065908 & 9.09$\times10^{-5}$ &  -                  \\
\hline
\noalign{\smallskip}
\end{tabular}
\centering
\end{table*}

\subsection{Partition function and equilibrium constants}

The internal partition function is given by a sum over all states:
\begin{equation}
Q_{int}=\sum_{e,v,N} (2N+1)(2S+1) \exp \left(\frac{-(E_{e,v,N}-E_0)}{kT}\right), \label{_eq1}
\end{equation}
where, $S$ is electronic spin,
$E_{e,v,N}$ is the energy of the state with electronic excitation $e$, 
vibrational excitation $v$ and with rotational angular momentum $N$, 
$E_0$ is the zero point energy. 
 When rotational angular momentum is coupled to the electronic 
spin a splitting of the spin degeneracy occurs. For the case of the CrH 
X~$^6\Sigma^+$ electronic states, $S$=5/2, this 
results in 6 separate states with angular momentum $J = N\pm$5/2, $N\pm$3/2, 
$N\pm$1/2. Similarly for MgH, X~$^2\Sigma^+$, $J = N\pm$1/2. 
The internal partition
function is then expressed as: 
\begin{equation}
Q_{int}=\sum_{e,v,J,N} (2J+1) \exp \left(\frac{-(E_{e,v,J,N}-E_0)}{kT}\right) \label{_eq2}
\end{equation}

Energy levels are computed using the rotation, vibration constants listed 
in Table
 \ref{_t2}. It is worth noting a good agreement of our 
data with the constants obtained by
Ram \& Bernath (1995) from the high resolution spectroscopy 
of the \ACrH - \XCrH
system of CrD molecule.
 The partition functions of MgH, MgD, CrD 
are computed via direct summation of these levels, see equation \ref{_eq1}. 
For CrH, we have obtained accurate energy levels 
from the linelist of Burrows et al. (2002). This list of energy levels
covers the $X~^6\Sigma^+$ and $A~^6\Sigma^+$ electronic states,
extends from $J$ = 0.5 to 39.5 and includes the vibrational states 
$v=$0,1,2,3 in the ground electronic state and $v=$0,1,2 for the 
excited electronic state. The CrH partition function is calculated
by the direct summation of the Burrows et al. (2002) energy levels. 
In addition, to cover the states omitted from the 
Burrows et al. (2002) linelist, we supplement energy levels 
computed with the spectroscopic constants. The partition functions 
are listed in Table \ref{_Q}.

\begin{table}
\centering \caption{\label{_Q} Internal partition functions, 
$Q$, as a function
of temperature (powers of ten are given in parenthesis).}
\begin{tabular}{rrrrr}
\hline
\noalign{\smallskip}
$T$ / K &  $Q$(MgH) & $Q$(MgD) & $Q$(CrH) & $Q$(CrD) \\
\hline
\noalign{\smallskip}

   50 & 1.322(1) & 2.440(1) & 3.608(1) & 6.855(1)\\
  100 & 2.575(1) & 4.815(1) & 7.011(1) & 1.351(2)\\
  150 & 3.832(1) & 7.194(1) & 1.042(2) & 2.018(2)\\
  200 & 5.091(1) & 9.582(1) & 1.384(2) & 2.687(2)\\
  250 & 6.353(1) & 1.199(2) & 1.726(2) & 3.359(2)\\
  300 & 7.623(1) & 1.445(2) & 2.069(2) & 4.041(2)\\
  350 & 8.905(1) & 1.698(2) & 2.415(2) & 4.738(2)\\
  400 & 1.021(2) & 1.961(2) & 2.764(2) & 5.455(2)\\
  450 & 1.154(2) & 2.236(2) & 3.118(2) & 6.199(2)\\
  500 & 1.290(2) & 2.524(2) & 3.479(2) & 6.973(2)\\
  550 & 1.431(2) & 2.825(2) & 3.849(2) & 7.780(2)\\
  600 & 1.577(2) & 3.142(2) & 4.228(2) & 8.623(2)\\
  650 & 1.727(2) & 3.474(2) & 4.618(2) & 9.503(2)\\
  700 & 1.883(2) & 3.823(2) & 5.020(2) & 1.042(3)\\
  750 & 2.045(2) & 4.188(2) & 5.435(2) & 1.138(3)\\
  800 & 2.212(2) & 4.569(2) & 5.864(2) & 1.238(3)\\
  850 & 2.386(2) & 4.969(2) & 6.306(2) & 1.343(3)\\
  900 & 2.566(2) & 5.385(2) & 6.763(2) & 1.451(3)\\
  950 & 2.753(2) & 5.820(2) & 7.235(2) & 1.564(3)\\
 1000 & 2.947(2) & 6.272(2) & 7.722(2) & 1.682(3)\\
 1100 & 3.355(2) & 7.233(2) & 8.744(2) & 1.930(3)\\
 1200 & 3.791(2) & 8.267(2) & 9.831(2) & 2.197(3)\\
 1300 & 4.256(2) & 9.378(2) & 1.098(3) & 2.482(3)\\
 1400 & 4.750(2) & 1.057(3) & 1.221(3) & 2.787(3)\\
 1500 & 5.274(2) & 1.183(3) & 1.350(3) & 3.111(3)\\
 1600 & 5.827(2) & 1.318(3) & 1.487(3) & 3.455(3)\\
 1700 & 6.410(2) & 1.460(3) & 1.630(3) & 3.818(3)\\
 1800 & 7.023(2) & 1.610(3) & 1.782(3) & 4.203(3)\\
 1900 & 7.664(2) & 1.767(3) & 1.941(3) & 4.608(3)\\
 2000 & 8.333(2) & 1.932(3) & 2.108(3) & 5.035(3)\\
 2100 & 9.029(2) & 2.104(3) & 2.283(3) & 5.483(3)\\
 2200 & 9.751(2) & 2.283(3) & 2.467(3) & 5.953(3)\\
 2300 & 1.050(3) & 2.469(3) & 2.659(3) & 6.446(3)\\
 2400 & 1.127(3) & 2.661(3) & 2.859(3) & 6.962(3)\\
 2500 & 1.206(3) & 2.859(3) & 3.069(3) & 7.501(3)\\
 2600 & 1.287(3) & 3.063(3) & 3.288(3) & 8.063(3)\\
 2700 & 1.370(3) & 3.272(3) & 3.517(3) & 8.650(3)\\
 2800 & 1.455(3) & 3.486(3) & 3.755(3) & 9.261(3)\\
 2900 & 1.542(3) & 3.705(3) & 4.003(3) & 9.896(3)\\
 3000 & 1.629(3) & 3.928(3) & 4.261(3) & 1.056(4)\\
 3100 & 1.719(3) & 4.154(3) & 4.530(3) & 1.124(4)\\
 3200 & 1.809(3) & 4.385(3) & 4.810(3) & 1.195(4)\\
 3300 & 1.901(3) & 4.618(3) & 5.101(3) & 1.269(4)\\
 3400 & 1.993(3) & 4.854(3) & 5.403(3) & 1.345(4)\\
 3500 & 2.086(3) & 5.093(3) & 5.716(3) & 1.424(4)\\
 3600 & 2.180(3) & 5.334(3) & 6.042(3) & 1.506(4)\\
 3700 & 2.275(3) & 5.577(3) & 6.379(3) & 1.590(4)\\
 3800 & 2.370(3) & 5.821(3) & 6.728(3) & 1.676(4)\\
 3900 & 2.465(3) & 6.067(3) & 7.089(3) & 1.765(4)\\
 4000 & 2.561(3) & 6.314(3) & 7.463(3) & 1.857(4)\\
 4100 & 2.657(3) & 6.562(3) & 7.849(3) & 1.951(4)\\
 4200 & 2.753(3) & 6.811(3) & 8.248(3) & 2.048(4)\\
 4300 & 2.849(3) & 7.060(3) & 8.659(3) & 2.148(4)\\
 4400 & 2.945(3) & 7.310(3) & 9.084(3) & 2.250(4)\\
 4500 & 3.041(3) & 7.559(3) & 9.521(3) & 2.354(4)\\
 4600 & 3.137(3) & 7.809(3) & 9.971(3) & 2.461(4)\\
 4700 & 3.233(3) & 8.059(3) & 1.043(4) & 2.571(4)\\
 4800 & 3.329(3) & 8.308(3) & 1.091(4) & 2.683(4)\\
 4900 & 3.424(3) & 8.557(3) & 1.140(4) & 2.797(4)\\
 5000 & 3.519(3) & 8.805(3) & 1.190(4) & 2.913(4)\\
\hline
\noalign{\smallskip}
\end{tabular}
\centering
\end{table}

\begin{figure*}
\begin{center}
\includegraphics [angle=-90,width=178mm]{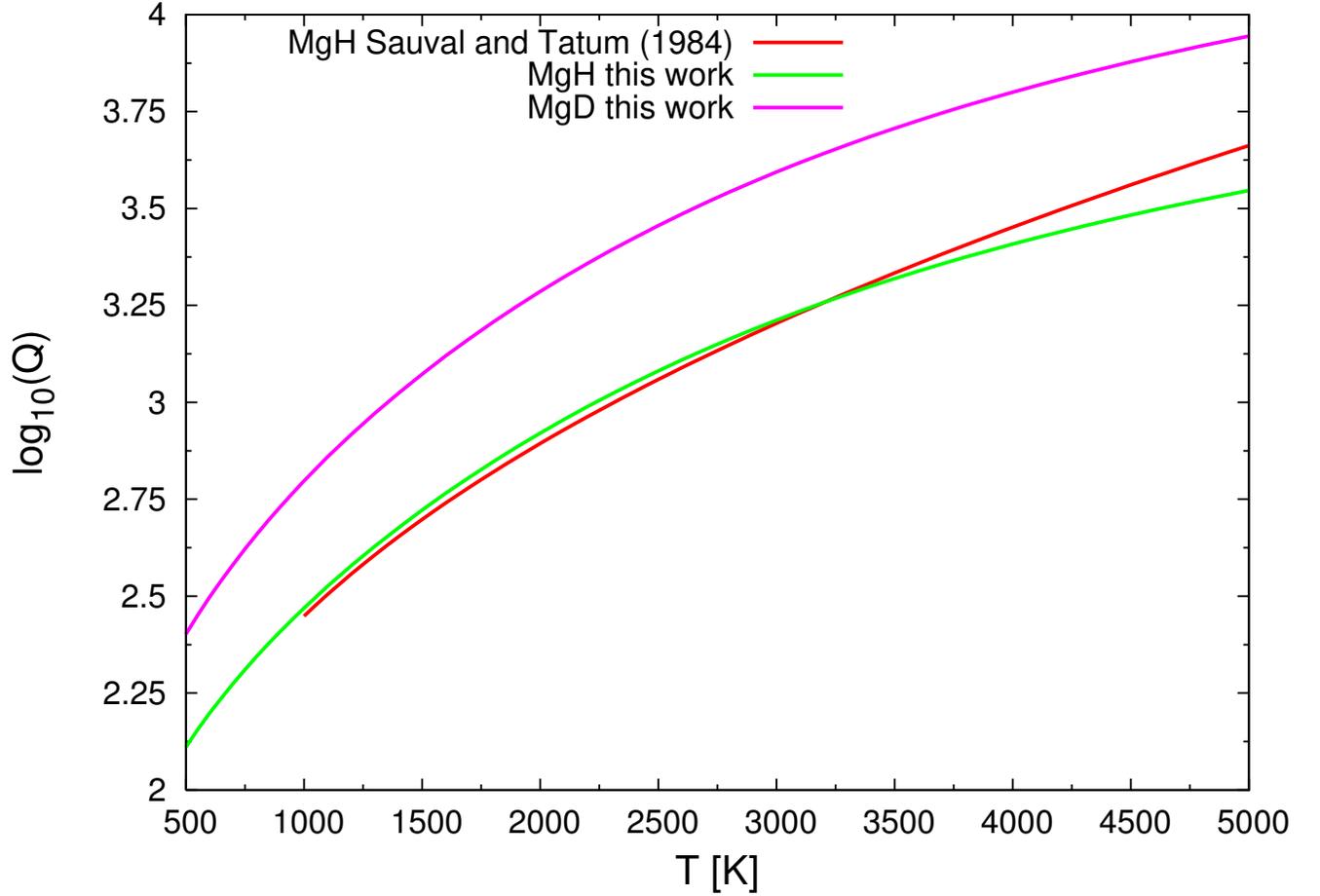}
\end{center}
\caption[]{\label{_figQMgH} Partition functions of MgH and MgD computed in 
this work, and the fit of Sauval \& Tatum (1984). }
\end{figure*}

\begin{figure*}
\begin{center}
\includegraphics [angle=-90,width=178mm]{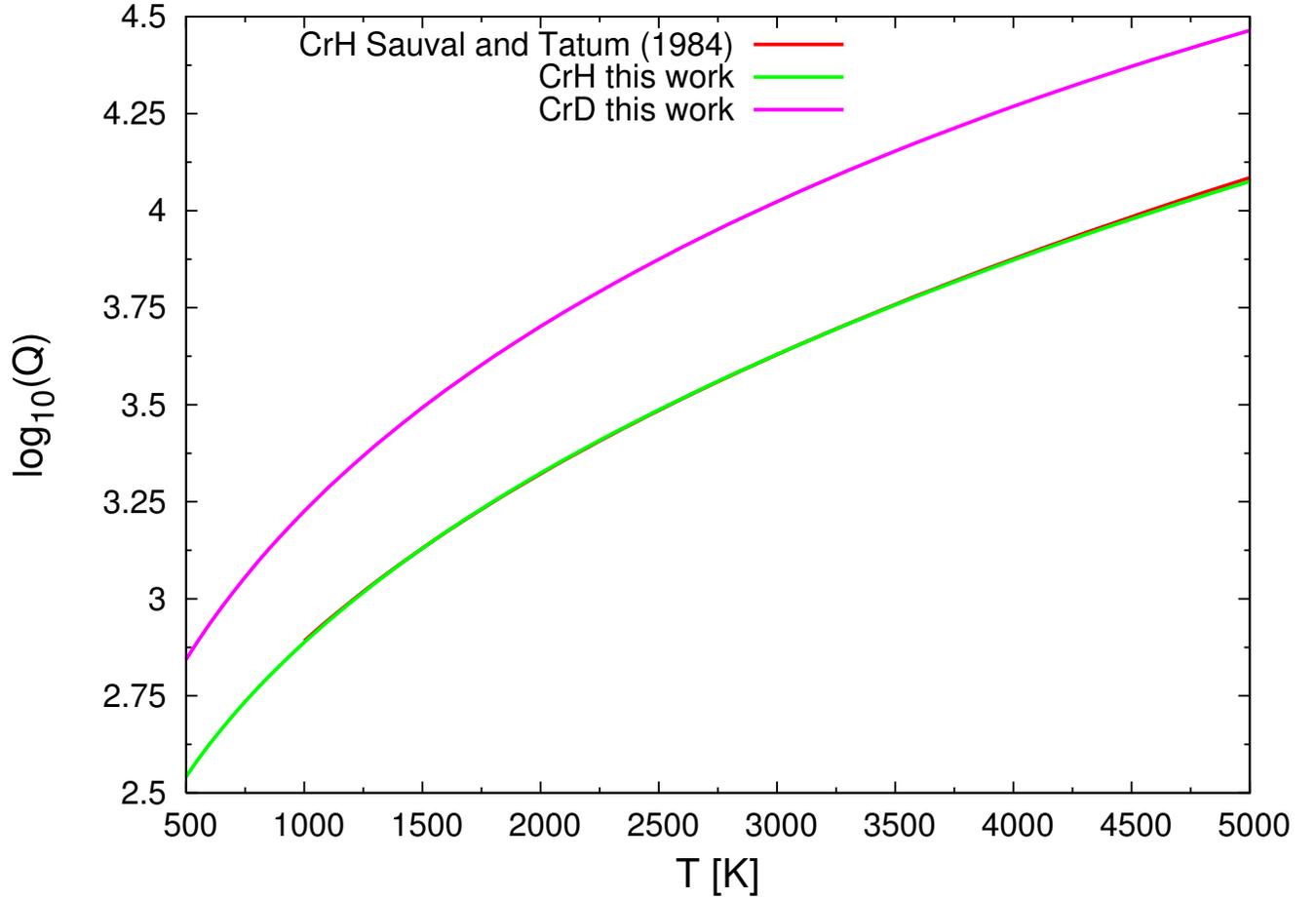}
\end{center}
\caption[]{\label{_figQCrH} Partition functions of CrH and CrD computed in 
this work, and the fit of of Sauval and Tatum (1984).
}
\end{figure*}

Figures \ref{_figQMgH} and \ref{_figQCrH} show the partition functions and the
respective fits of Sauval \&  Tatum (1984). For CrH there is
excellent agreement with the partition function of Sauval \&  Tatum (1984), but for 
MgH our partition function is slightly larger than that of Sauval \&  Tatum (1984) 
at low temperature and smaller at temperatures above 3200~K.
Sauval \&  Tatum (1984) make use of a high temperature
approximation to the rotational partition function. This uses the rigid rotor
approximation and is determined by rewriting the
partition function sum as an integral from zero to infinity. In contrast our partition function
sum is truncated at the dissociation energy D$_0$, which for MgH is only 1.285 eV. Thus at
high temperature Sauval \&  Tatum (1984) predict a higher partition function than would
be expected by direct summation of bound states.

We have calculated equilibrium constants for the reactions: 
MgH$+$D$\rightleftharpoons$MgD$+$H 
and CrH$+$D$\rightleftharpoons$CrD$+$H. The dimensionless equilibrium constants are given by:
\begin{eqnarray}
K & = & \frac{n(CrH)n(D)}{n(CrD)n(H)} \nonumber \\
  & = & \frac{Q(CrH)Q(D)}{Q(CrD)Q(H)}\exp\left(-\frac{\Delta E}{kT}\right)
\end{eqnarray}
where $n$ are the number densities,  and $\Delta E$ is 
the energy difference between zero point energies of CrH and CrD;
$Q$ is the total partition functions which can be factorised into internal 
and translational
components, $Q=Q_{int}Q_{tran}$, where $Q_{tran}=(2\pi M_ikT/h^2)^{3/2}$.
$g_j$ is the degeneracy of energy level $j$. By convention nuclear spin 
degeneracy is 
omitted but electronic spin degeneracy is included. Thus for a 
proton $Q_{int} =1$ 
and for an electron $Q_{int} = 2$.
The equilibrium constants 
are listed in Table \ref{_eqlib}.
The equilibrium 
constants indicate that at low temperatures the formation of the deuterides are 
strongly favoured over the hydrides, but at high temperatures the hydrides become weakly favoured.

 It is worth noting that in cool dwarfs hydrogen and deuterium are primarily
 found in their molecular forms, H$_2$, HD and D$_2$. 
In a full chemical equilibrium calculation
the partition functions and dissociation energies of H$_2$, HD and D$_2$ will
have an affect upon the abundances of the molecules of interest.
So a more appropriate reaction may be MgH + HD $\rightleftharpoons$
   MgD + H$_2$. 
However, to calculate equilibrium constant for this reaction would require 
the partition functions of Q(HD) and Q(H$_2$) and their dissociation energies.
Expressing the equilibrium constants relative to atomic D and H
avoids any uncertainties arising from the partition functions and 
dissociation energies of H$_2$, HD and D$_2$. A calculation of the partition 
functions and dissociation energies of H$_2$, HD and D$_2$ is beyond the 
scope of this project.

In calculating the equilibrium constants we
assumed that the electronic partition functions of H and D are identical.
This is a good approximation as the main differences in the
energy levels are proportional to the change in reduced mass and are thus small.


\begin{table}
\centering \caption{\label{_eqlib} Equilibrium constants for deuterium
exchange reactions (powers of ten are given in parenthesis).}
\begin{tabular}{rll}
\hline
\noalign{\smallskip}
 $T$/K &  $K$(MgH+D$\rightleftharpoons$MgD+H) & $K$(CrH+D$\rightleftharpoons$CrD+H)\\
\hline
\noalign{\smallskip}

   50 & 3.998(3) & 1.759(3)\\
  100 & 7.504(2) & 4.973(2)\\
  150 & 1.994(1) & 1.515(1)\\
  200 & 3.250(1) & 2.644(1)\\
  250 & 4.350(1) & 3.690(1)\\
  300 & 5.270(1) & 4.600(1)\\
  350 & 6.028(1) & 5.372(1)\\
  400 & 6.647(1) & 6.020(1)\\
  450 & 7.152(1) & 6.561(1)\\
  500 & 7.567(1) & 7.012(1)\\
  550 & 7.908(1) & 7.390(1)\\
  600 & 8.191(1) & 7.707(1)\\
  650 & 8.427(1) & 7.975(1)\\
  700 & 8.625(1) & 8.201(1)\\
  750 & 8.793(1) & 8.395(1)\\
  800 & 8.935(1) & 8.560(1)\\
  850 & 9.057(1) & 8.703(1)\\
  900 & 9.162(1) & 8.826(1)\\
  950 & 9.254(1) & 8.933(1)\\
 1000 & 9.333(1) & 9.027(1)\\
 1100 & 9.464(1) & 9.182(1)\\
 1200 & 9.567(1) & 9.304(1)\\
 1300 & 9.648(1) & 9.400(1)\\
 1400 & 9.714(1) & 9.479(1)\\
 1500 & 9.768(1) & 9.542(1)\\
 1600 & 9.812(1) & 9.595(1)\\
 1700 & 9.849(1) & 9.640(1)\\
 1800 & 9.880(1) & 9.677(1)\\
 1900 & 9.906(1) & 9.709(1)\\
 2000 & 9.928(1) & 9.737(1)\\
 2100 & 9.947(1) & 9.761(1)\\
 2200 & 9.963(1) & 9.783(1)\\
 2300 & 9.977(1) & 9.803(1)\\
 2400 & 9.988(1) & 9.822(1)\\
 2500 & 9.998(1) & 9.840(1)\\
 2600 & 1.001 & 9.857(1)\\
 2700 & 1.001 & 9.874(1)\\
 2800 & 1.002 & 9.891(1)\\
 2900 & 1.003 & 9.909(1)\\
 3000 & 1.003 & 9.927(1)\\
 3100 & 1.003 & 9.945(1)\\
 3200 & 1.004 & 9.965(1)\\
 3300 & 1.004 & 9.985(1)\\
 3400 & 1.004 & 1.001\\
 3500 & 1.004 & 1.003\\
 3600 & 1.004 & 1.005\\
 3700 & 1.005 & 1.008\\
 3800 & 1.005 & 1.011\\
 3900 & 1.005 & 1.013\\
 4000 & 1.005 & 1.016\\
 4100 & 1.005 & 1.019\\
 4200 & 1.005 & 1.022\\
 4300 & 1.005 & 1.026\\
 4400 & 1.005 & 1.029\\
 4500 & 1.005 & 1.032\\
 4600 & 1.005 & 1.036\\
 4700 & 1.005 & 1.039\\
 4800 & 1.005 & 1.043\\
 4900 & 1.005 & 1.047\\
 5000 & 1.005 & 1.051\\

\hline
\noalign{\smallskip}
\end{tabular}
\centering
\end{table}

\subsection{Ionisation-dissociation equilibrium}

The equations of ionisation-dissociation equilibrium were solved for media
consisting of atoms, ions and molecules. 
We took into account
$\sim$ 100 components (Pavlenko 1998, 2000). The constants for equations
of chemical balance were taken from Tsuji (1973)
and Gurvitz et al. (1989).
To find the densities of species we apply the technique applied 
by Kurucz (1975) in the ATLAS5 -- ATLAS12 programs. In the framework of
the local thermodynamic equilibrium (LTE) approach,  the densities of a 
molecule consisting of atoms labelled $1, 2, ... n$ in the $k+$ ionisation stage
can be described by the Gouldberg-Waage equation:
\begin{eqnarray}
\frac{n_{1,2,...l}^{p^+}}{\prod_i^l n_i}= \frac{Q_{1,2...n} Q_e^p}{\prod_i^n Q_i} \exp\left(-\frac{E_{1,2,... l}}{kT}\right), \label{eq1} 
\end{eqnarray}
 where $n_i$ is the number density of the $i^{\rm th}$ component, $Q_i$ are 
the total partition functions for species i, and $E_{1,2,...l}$ is the 
dissociation energy (D$_0$) of molecule which consists of $l$ atoms and has $p^+$ charge. 

In the ATLAS program a computationally more convenient format is used:

\begin{eqnarray}
\frac{\prod_i ^l N_i }{N_{1,2,...l}} & = &\exp[-E_{1,2,... l}/kT + g(T)],
\label{eq3}
\end{eqnarray}
here the function $g(T)$ is given by:

\begin{eqnarray}
g(T) & = & b-c(T+d(T-e(T+fT)))\\ \nonumber
 & & +\frac{3}{2}(l-p-1)\ln T,
\end{eqnarray}
values of $b,c,d,e$ (Table \ref{abcd}) are determined by the 
fitting to the data (see Table \ref{_eqlib})  using the least square 
minimisation procedure for the temperature range 400 - 5000 K.

As was expected, at LTE the equilibrium constants of our hydrides and deuterides
are very similar. Indeed, the equilibrium constants 
are $\sim$ $Q_{int, i}/M_i$, $M_i$ is the mass of the $i^{\rm th}$ species.
Partition functions of the deuterated molecules 
are larger by approximately a factor 2, which corresponds to the 
larger molecular weight of deuterium. Fitted chemical equilibrium constants
computed for CrH, CrD, MgD and MgH are shown in Table \ref{abcd}.

\begin{table*}
\caption{\label{abcd}Chemical equilibrium constants for MgH, MgD, CrH and CrD (powers of ten in parenthesis).}
\begin{tabular}{lllllll}
\hline
\hline
\noalign{\smallskip}
Molecule& $D_{0}$ /eV &    $b$    &   $c$      &    $d$     &    $e$    &   $f$       \\
\noalign{\smallskip}
\hline
\noalign{\smallskip}
          &       &            &            &            &            &            \\
CrH & 2.863 & 3.690(1) & 6.678(-3) & 2.541(-6) & 5.115(-10) & 3.929(-14) \\
MgH & 1.285 & 3.605(1) & 6.834(-3) & 2.551(-6) & 5.049(-10) & 3.834(-14) \\
          &       &            &            &            &            &            \\
CrD & 2.888 & 3.731(1) & 6.919(-3) & 2.607(-6) & 5.192(-10) & 3.967(-14) \\
MgD & 1.311 & 3.648(1) & 7.111(-3) & 2.647(-6) & 5.214(-10) & 3.942(-14) \\
          &       &            &            &            &            &            \\
\hline

\multicolumn{7}{l}{ \footnotesize $^a$ From Shayesteh et al. (in prep.) } \\

\end{tabular}

\end{table*}

\subsection{Synthetic spectra}

Synthetic spectra were computed using the program
WITA6 (Pavlenko 2000) assuming LTE, hydrostatic equilibrium and a
one-dimensional model atmosphere without sources and sinks of
energy. Hereafter we use the term ``synthetic spectra'' instead of
``spectral energy distributions'' to simplify the text.
Theoretical synthetic spectra were computed for the
model atmospheres of dwarfs with effective temperatures T$_{\rm
eff}$ = 1200 -- 1800 K from the COND grid of Allard at al.(2001) using
solar metallicity (Anders \& Grevesse 1989). Unless otherwise
mentioned all models are for log g = 5.0.

\subsection{Atomic lines opacity}

To compute the band profiles of the electronic bands in the spectra of late
spectral class dwarfs one must account for absorption of the resonance
lines of Na I (0.5891, 0.5897 \mum) and  K I (0.7667, 0.7701
\mum), which are extremely strong. In fact, optical and near
infrared spectra of L-and T-dwarfs are governed by them (Pavlenko
1997, 1998). Temperatures in the atmospheres of ultracool dwarfs are
lower than 2000 K, therefore alkali atoms exist there mainly in the
form of neutrals. Furthermore, due to the low opacity of the 
atmospheres of ultracool dwarfs their photospheres occur in
 high pressure layers. Strong resonance lines of the most
abundant alkali metals, i.e sodium and potassium, are formed in
very dense and cool  plasmas. Due to the high efficiency of pressure
broadening, their formally computed equivalent widths may reach a
few thousand Angstroms (Pavlenko 2000). Burrows \& Volobuev (2003) show
that we cannot use the conventional theory of collisional
broadening. A more sophisticated theory involving quasi-stationary
broadening should be used for them (see also Allard et al. 2003,
and Zhu et al. 2006).

In this work we use the potentials of quasi-stationary chemical
interactions of K and Na with the most numerous species,
 atomic He and molecular H$_2$ computed by GAMESS (Granovsky et
al. 2004). Our procedure is described in more detail in Pavlenko
et al. (2007). In calculations of K I
profiles we used a combined profile: the cores of these lines were
computed in the framework of the collisional approach and their wings
($\delta\lambda >$ 0.004 \mum) were treated by quasi-stationary
theory. 

\section{Results}

\subsection{MgH \& MgD}

\begin{figure*}
\begin{center}
\includegraphics [width=178mm]{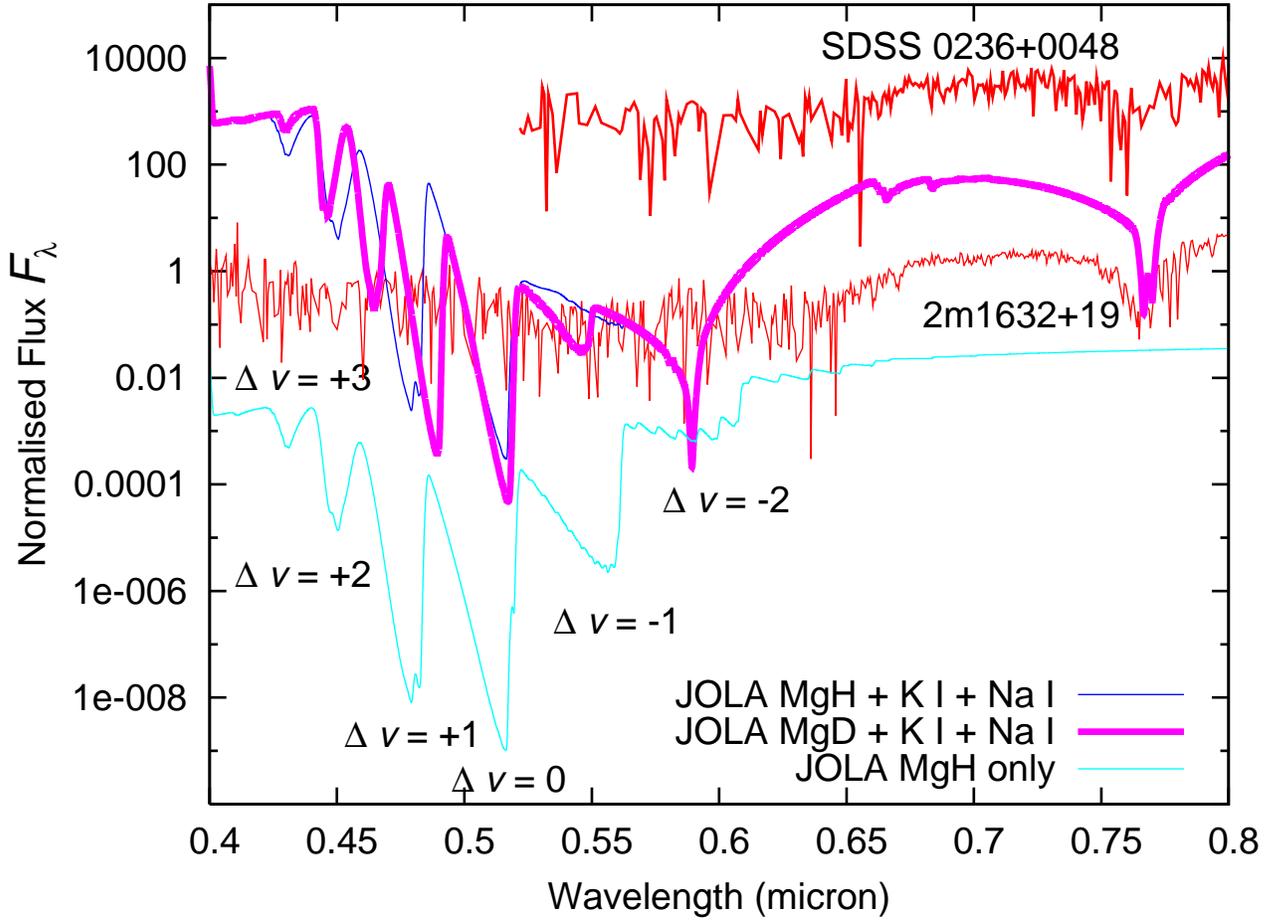}
\end{center}
\caption[]{\label{_MgH} Molecular bands of the \AMgH --- \XMgH  system
of MgH and MgD in the theoretical spectrum computed for a 1800/5.0/0
COND model atmosphere. Spectra are shown for D/H = 1.
The lower line on the plot shows the pure
spectrum of MgH computed in the framework of JOLA approximation.
For comparison the observed spectra of
two L dwarfs 2MASS1632+19 (\Martin et al. 1999) and SDSS 0236+0048
(Leggett et al. 2001) are shown as solid lines. 
}
\end{figure*}

Computed JOLA profiles of the electronic band system \AMgH -- \XMgH of  MgH
and MgD are shown in Fig. \ref{_MgH}. It is worth noting that
a simulation based on a sound                                                   
quantum mechanical model, including spin-orbit coupling in the \AMgH state,       
 agreed reasonably well with the JOLA calculation (see Hill 2007).    
First, the synthetic spectra were computed for the case of a 
solar abundances (N(H)/N(total)= 0.9). Then
all the hydrogen was replaced by deuterium (N(D)/N(Total)=0.9).
As stated above, the aim of this paper is a comparative
analyses of the possibilities of detection of the deuterated
molecules MgD and CrD which are formed on the background of the
electronic bands of the more abundant conventional hydrides MgH and CrH.
The detection of MgD bands is very challenging for several reasons:

\begin{itemize}
\item The bands of MgD lie at relatively short wavelengths. High precision
observations of this spectral region are difficult 
in objects with low \Tef . 

\item The differences in the positions of the band heads of MgH and MgD are
rather small. The MgD bands will be formed on a
background of far stronger MgH bands.

\item MgH and MgD bands in the spectra of cool dwarfs 
are close to a strong Na I resonance line and the lines of 
other molecules/atoms.
\end{itemize}

\subsection{CrH \& CrD}

\begin{figure*}
\begin{center}
\includegraphics [width=178mm]{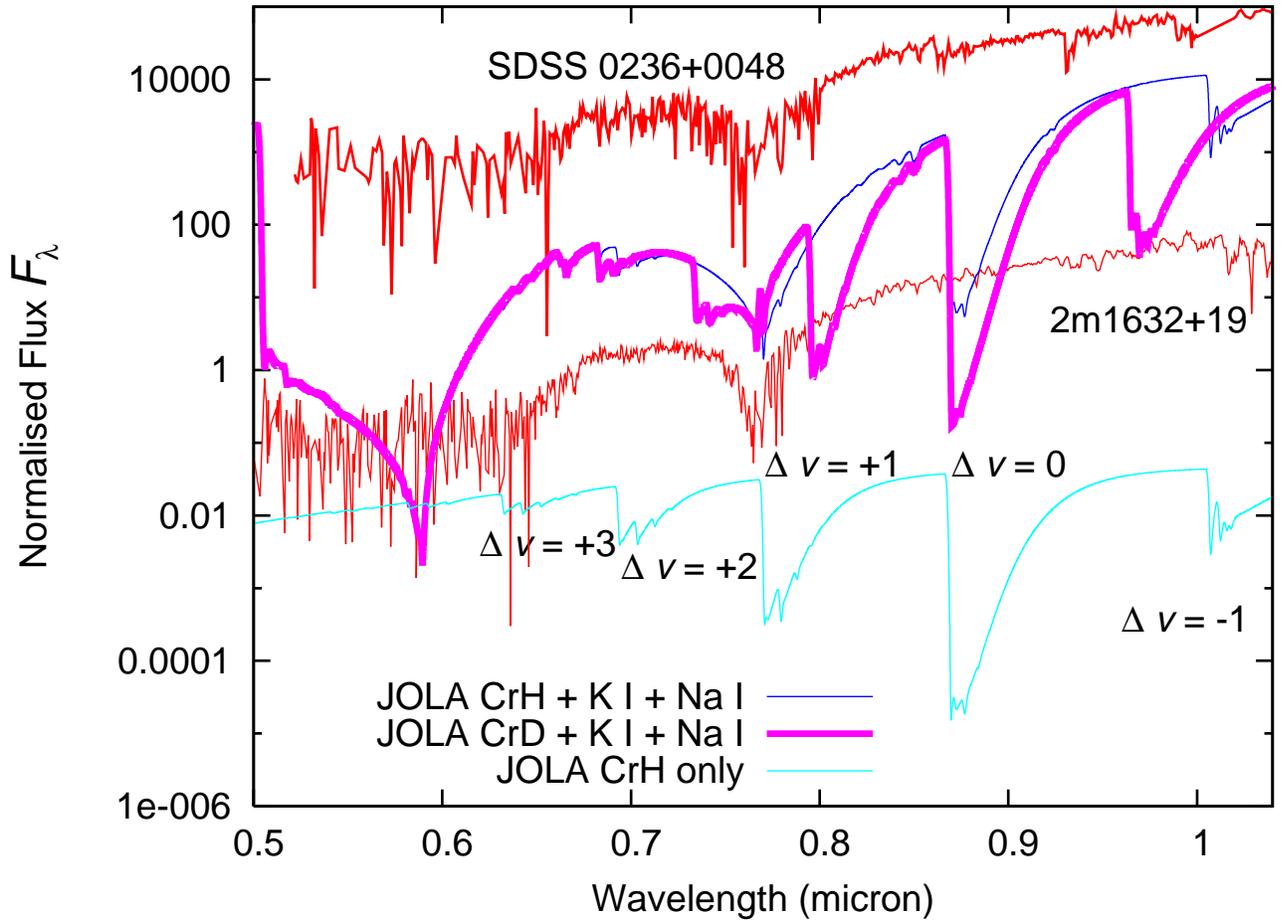}
\end{center}
\caption[]{\label{_CrH} Molecular bands of the \ACrH --- \XCrH  system 
of CrH and CrD and a theoretical spectrum computed for a 1800/5.0/0 
COND model atmosphere. Spectra are shown for D/H =1.
The lower line on the plot shows the pure 
spectrum of MgH computed in the framework of JOLA approximation.
For comparison the observed spectra of 
two L dwarfs 2MASS1632+19 (\Martin et al. 1999) and SDSS 0236+0048 
(Leggett et al. 2001) are shown as solid lines. 
} 
\end{figure*}

The relative behaviour of CrH and CrD bands is shown in Fig.
\ref{_CrH}. In general, the bands of the \ACrH -- \XCrH system occupy
a  larger wavelength range.
Due to the larger reduced mass, distances between heads of CrD
bands of different $\Delta v$ are smaller  than those for the CrH. As a
result, bands of negative and positive $\Delta v$ are shifted
bluewards  and redwards with respect to the corresponding CrH bands.

The transitions with $\Delta v$ = 1 form a band head at 0.76843 \mum
which coincides with the core of the strong K I resonance doublet.
Thus in L dwarfs with strong CrH absorption the
contribution of CrH has to be accounted in the modelling of 0.77
\mum K I feature.

For the $\Delta v=0$ case the bands of CrH and CrD are co-incident.
For the $\Delta v$ positive case the CrD band heads are displaced 
to longer wavelengths of the CrH band head, thus the strongest CrD lines will
be blended with the lines from the tail of the corresponding CrH band.
Bands with $\Delta v$ = 2 are blended with the K I
resonance line. Bands with $\Delta v$ = 3 are located in the
spectral region of the so-called ``satellite rainbows'' of the K I
line (Burrows \& Volobuev 2003).
For the $\Delta v$ negative case the CrD band heads are displaced 
to shorterer wavelengths of the CrH band head, these bands offer the best possibility of
detection.

\section{Discussion}

The detection of deuterated molecules in the spectra of ultracool
dwarfs provide a challenge for both theoreticians and
observers. Indeed, in the atmospheres of planetary mass objects
(M $<$ 13 M$_J$) we cannot expect ratio D/H $> 2\times10^{-5}$. This
means the lines of deuterated molecules should be about 5000 times
weaker than those of the hydrides. The ideal case would be a spectral 
region where the molecular bands are not blended. 
So a crucial requirement is a large difference in the wavelengths of the
band heads of the hydrides and deuterides.

CrH appears to be more useful than MgH in the search for deuterated species.
The band heads of CrD are displaced significantly from the band heads of
CrH. Bands of CrH are observed in the spectra of the latest L
dwarfs (Kirkpatrick et al. 1999). The CrD bands are located 
in the ``near infrared'' spectra, where
fluxes are much higher than in the ``optical'' spectral
regions.

In this paper we show that the most useful bands for the 
realisation of the deuterium test are $\Delta v = +1$  and
$\Delta v = -1$ ($\lambda$ = 0.795 and 0.968 \mum,
respectively). The $\Delta v = -1$ band looks especially promising.
It is located in the near infrared region with an absence of strong background
absorption features.
 However, portions of this CrD band will                                         
be swamped by the 0-0 FeH Wing-Ford band at 0.99$\mu$m and possibly by          
water bands. 
Still, the case for CrD looks better than for HDO lines which are      
formed on a                                                                     
background of strong H2O lines (Pavlenko 2002). High quality line lists         
are required to test these possibilities fully.

It is worth noting, that a potential problem lies in the possibility that Cr atoms are absorbed 
onto dust particles. The
depletion of Cr will reduce the strength of both CrD and CrH bands. 
Fortunately as CrH bands are
located in the same spectral region, we can ``scale'' the CrD depletion processes
by adjusting the strength of CrH bands.

For more precise studies, more accurate and 
detailed linelists of CrH and CrD are required. The calculation of such linelists
would require new improved computations supported by new laboratory
measurements. 
%
One problem that concerns us is that even once we have a good agreement           
with the model and experimental data for the MgH or CrH molecule, there         
are still perturbations about which we know very little from experiments         
done so far. Nonetheless, MgH is a non-starter for the
deuterium test and as we note above, the model adopted in our paper is more likely to be reliable for CrH and CrD.
%
It is worth noting that the use of the pure rotational-vibrational bands 
located in the mid and far infrared spectral region may offer an
alternative. Indeed, the displacement between CrH and CrD rotation-vibration bands 
is even larger, than for the case of electronic bands. 


Nevertheless, we cannot be certain that we have identified the 
best candidate systems
for the deuterium test. 
Future investigations of deuterated molecules
in different spectral regions are important to determine which offers the best 
possibility for the realisation of this test. 
The ideal solution would be to detect lines of deuterated molecule(s)
in different spectral regions. This presents a serious 
observational challenge which can only be met in combination with careful
laboratory measurement and the computation of high quality molecular spectra.

\section{Acknowledgements}

We thank anonymous Referee for the helpful remarks. We thank Prof. Peter Bernath
for a constructive discussion and for providing us with a copy of his paper
prior to publication.
The UK Particle Physics and Astronomy research council is thanked for
visitor and PDRA support. The Royal Society is thanked for a visitor grant.
YP is supported by a Leverhulme Trust grant. 
This work was partially
supported by the Cosmomicrophysics program of National Academy of Sciences
and National Space Agency of Ukraine.
Our research has made
use of the SIMBAD database operated at CDS, Strasbourg, France.


\label{lastpage}
\end{document}